\documentclass[final]{svjour3}
\pdfoutput=1
\usepackage{gensymb}
\usepackage{graphicx}
\usepackage{rotating}
\usepackage{amssymb}
\usepackage{mathptmx}
\usepackage{authblk}
\usepackage{sidecap}
\usepackage[numbers]{natbib}

\newcommand{\jcap}{\ref@jnl{J. Cosmology Astropart. Phys.}}%

\makeatletter
\journalname{Journal of Low Temperature Physics}

\bibpunct{}{}{,}{s}{}{,}
\begin{document}
\titlerunning{Superconducting On-chip Fourier Transform Spectrometer}
\authorrunning{R. Basu Thakur. N. Klimovich, P. K. Day, E. Shirokoff, et. al.}
\title{Superconducting On-chip Fourier Transform Spectrometer}

\author[1, 2]{R. Basu Thakur* \thanks{*ritoban@caltech.edu}}
\author[1, 2]{N. Klimovich}
\author[2]{P. K. Day}
\author[3, 4]{E. Shirokoff}
\author[5]{P. D. Mauskopf}
\author[5]{F. Faramarzi}
\author[6]{P. S. Barry}

\affil[1]{Department of Physics, California Institute of Technology, Pasadena, California 91125, USA}
\affil[2]{Jet Propulsion Laboratory (NASA), 4800 Oak Grove Dr., Pasadena, California 91109, USA}
\affil[3]{Kavli Institute for Cosmological Physics, University of Chicago, 5640 S. Ellis Ave., Chicago, IL,60637, USA}
\affil[4]{Department of Astronomy and Astrophysics, University of Chicago, 5640 S. Ellis Ave., Chicago, IL, 60637, USA}
\affil[5]{Arizona State University, Tempe, AZ 85281 USA}
\affil[6]{High-Energy Physics Division, Argonne National Laboratory, 9700 South Cass Avenue., Argonne, IL, 60439, USA}

\maketitle
\begin{abstract}
Kinetic inductance in thin film superconductors has been used as the basis for low-temperature, low-noise photon detectors. In particular thin films such as NbTiN, TiN, NbN, the kinetic inductance effect is strongly non-linear in the applied current, which can be utilized to realize novel devices. We present results from transmission lines made with these materials, where DC (current) control is used to modulate the phase velocity thereby enabling an on-chip spectrometer. The utility of such compact spectrometers are discussed, along with their natural connection with parametric amplifiers. 

\end{abstract}
\section{Introduction}
\begin{figure}[h!!]
  \centering
    \includegraphics[width=0.45\textwidth]{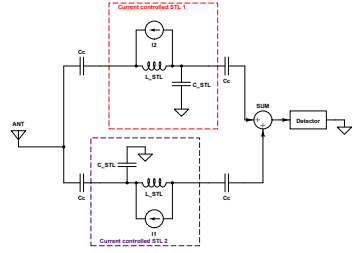}
  \caption{\small{SOFTS schematic: each STL is modeled as current controlled LC circuit (dashed boxes). STLs are capacitatively coupled~(Cc) to an antenna (ANT, left) receiving broadband power, and a summing junction (SUM, right) where the phase delayed signals are added and sensed by a detector (TES /MKID).}}
  \label{fig:softs2}
\end{figure}

In thin-film superconductors inductance is the sum of a constant value from device geometry ($L_g$), and a kinetic component ($L_k$) due to quasiparticle motion. $L_k$ nonlinearly depends on the super-current~\cite{Parmenter, Prober} ($I$), and for superconducting resonators / transmission lines the resonance frequency / transmission speed can be adjusted with a DC current \cite{DayNature, Kher, Che, STLLK, Ed}. The total inductance is expressed as $\displaystyle L ( I ) \approx (L_g + L_{k,0})\left(1+ \alpha\left(I/I_*\right)^2 + \alpha'(I/I_*)^4\right)$, where $I_*$ is a current-scale for non-linearity, $\alpha < 1$ is the canonical kinetic inductance fraction and $\alpha'$ is material and geometry dependant constant. The phase-velocity ($u$) in a superconducting transmission line (STL) can thus be current-controlled, see Eqn.~\ref{eq:uI}, where  the per-unit-length inductance/capacitance are $\mathcal{L}$/$\mathcal{C}$ respectively. Note that $\mathcal{C}$ and $ \mathcal{L}_g$ (geometric) can be designed during fabrication, while kinetic inductance is DC tunable. Therefore phase-velocity is engineer-able and controllable around desired values.

\begin{equation}
	u(I) = 1/ \sqrt{ \mathcal{C} \left( \mathcal{L}_g +  \mathcal{L}_{k,0}\right) \left( 1+ \alpha (I/I_*)^2 + \alpha'(I/I_*)^4 \right)  }
\label{eq:uI}
\end{equation}
\\Utilizing phase velocity control, we demonstrate a  Superconducting On-chip Fourier Transform Spectrometer (SOFTS). Two STLs are fed with identical inputs obtained from a source via a splitter. Each STL is one arm of a Mach-Zhender interferometer and the phase velocity/delay, is independently controlled. The interferogram obtained from the summing the STLs' outputs gives the source-spectrum via a Fourier transform, Fig.~\ref{fig:softs2}. For currents $I_1=0$ and $I_2=I$, the phase delay and velocity are related as $\Delta \phi(I) = 2\pi \nu x \left(u(I)^{-1}-u(0)^{-1} \right) $, here the photon frequency is $\nu$, and STL physical lengths are $x$.

\vspace{-0.2cm}
Such devices can be easily arrayed into wide-band \textbf{imaging$\times$spectral} kilopixel Integral Field Unit (IFU). A common IFU can serve a wide variety of science cases, especially with bandwidth and resolution being engineer-able. This new technology can replace meter-scale opto-mechanical FTSs and grating spectrometers with cm-scale electronic devices. Such unique IFUs enable a new class of instruments for measuring CMB spectral distortions which probes cosmology from pre-recombination epoch, studying physics from the epoch of reionization via line intensity mapping, as well as individual astrophysical objects like galaxy-clusters (SZ effect). Additionally such IFUs readily extend capabilities to map localized dusty regions, vital to understanding galaxy and star formation and for foreground cleaning for probes of inflation such as PICO and CMB-S4~\cite{PICO2019}, \cite{S4tech},\cite{SerraDore}.
\vspace{-0.2cm}
\section{STL design and fabrication}
\begin{figure}[!h]
	\centering
	\begin{minipage}[t]{4cm}
		\centering
		\includegraphics[scale=0.25]{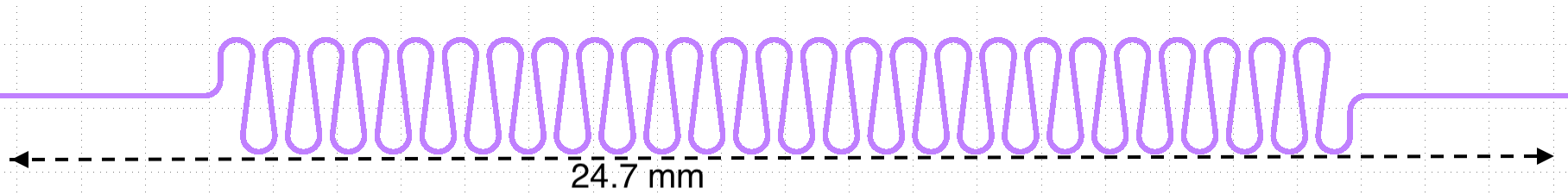}
		\caption{A ``zipper'' STL: 115mm total length, on-chip length 24.7mm.}
		\label{fig:zipper1}
	\end{minipage}
	\hspace{3.5cm}
	\begin{minipage}[t]{4cm}
		\centering
		\includegraphics[scale=0.25]{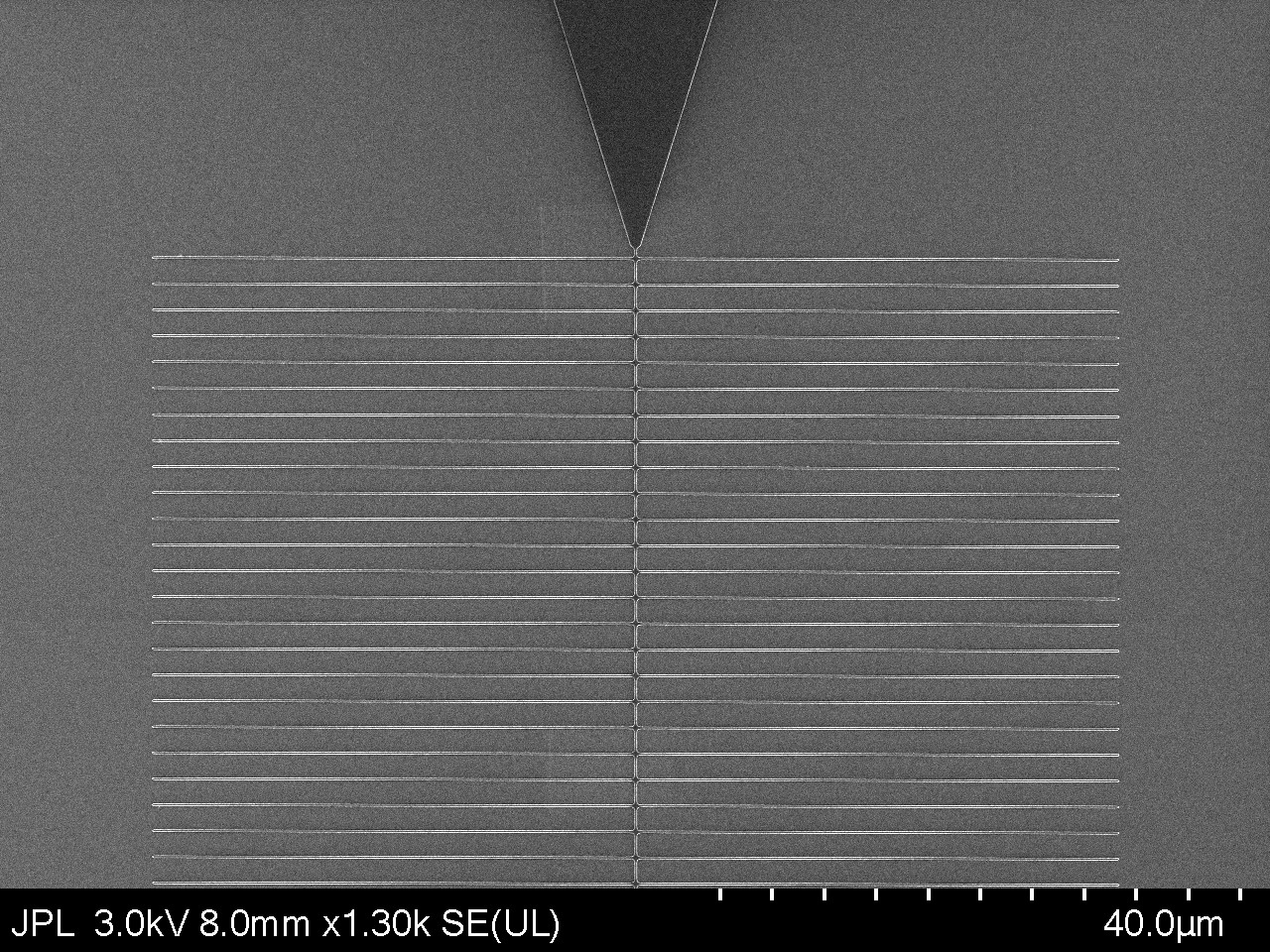}
		\caption{Micrograph (zoom in) of STL near a bondpad.}
		\label{fig:zipper2}
	\end{minipage}
\end{figure}

\vspace{-0.2cm}
The STLs used were originally designed for wideband, high dynamic range, low-noise superconducting amplifiers,~\cite{DayNature}. These are microstrip transmission lines, each device fabricated with 35nm layer of NbTiN on top of a crystalline Si substrate. Above the transmission line is 190nm of amorphous Si dielectric that separates it from the top layer of a 50nm NbTiN ground plane. The transmission line is 250nm wide and has periodic 250nm wide fingers for added capacitance and impedance matching, Fig.~\ref{fig:zipper1}. These thin-films have a superconducting transition at $T_c \approx 15$K, therefore in application to sub-mm science we have significantly broad bandwidths ($\lesssim 1$THz). Loss at the high frequencies will determine the optimal STL design, a topic of active research beyond the scope of this article. However we note several sub-mm projects have devices operating over 200 GHz~\cite{SPIE3G}, ~\cite{SuperSpec}. 
\vspace{-0.1cm}
\vspace{-1cm}
\section{STL performance}
\subsection{Phase delay measurements}
The crux of both parametric amplification and interferometric operations is sufficient slow down of light-speed and therefore controlled phase delay. Our measurements and fit model are shown in Fig.~\ref{fig:dphidf} and Eqn.~\ref{eq:dphidf_model}. The model has two parameters ($a_I, b_I$) and the coefficient $K$ is extracted from data by fitting the phase at zero current with frequency. The summary of our measurements are presented in table~\ref{tab:currvals}. 
All parametric errors $< 3 \%$. Since these STLs have $\alpha \rightarrow 1$, we may infer $I_* = a_I =$ 3mA, which implies $\alpha' =$3.16.

\begin{center}
\begin{tabular}{ |c|c|c|c|c| } 
 \hline
 K (radians/GHz) & $u(0)/c \%$ & $I_c$ (mA) & $a_I$ (mA) & $b_I$ (mA) \\ 
 \hline
 319.4 & 0.75 & 0.86 & 3.00 & 2.25 \\ 
 \hline
\end{tabular}
\label{tab:currvals}
\end{center}

\begin{minipage}{.5\textwidth}
    \centering
    \includegraphics[width=1\textwidth]{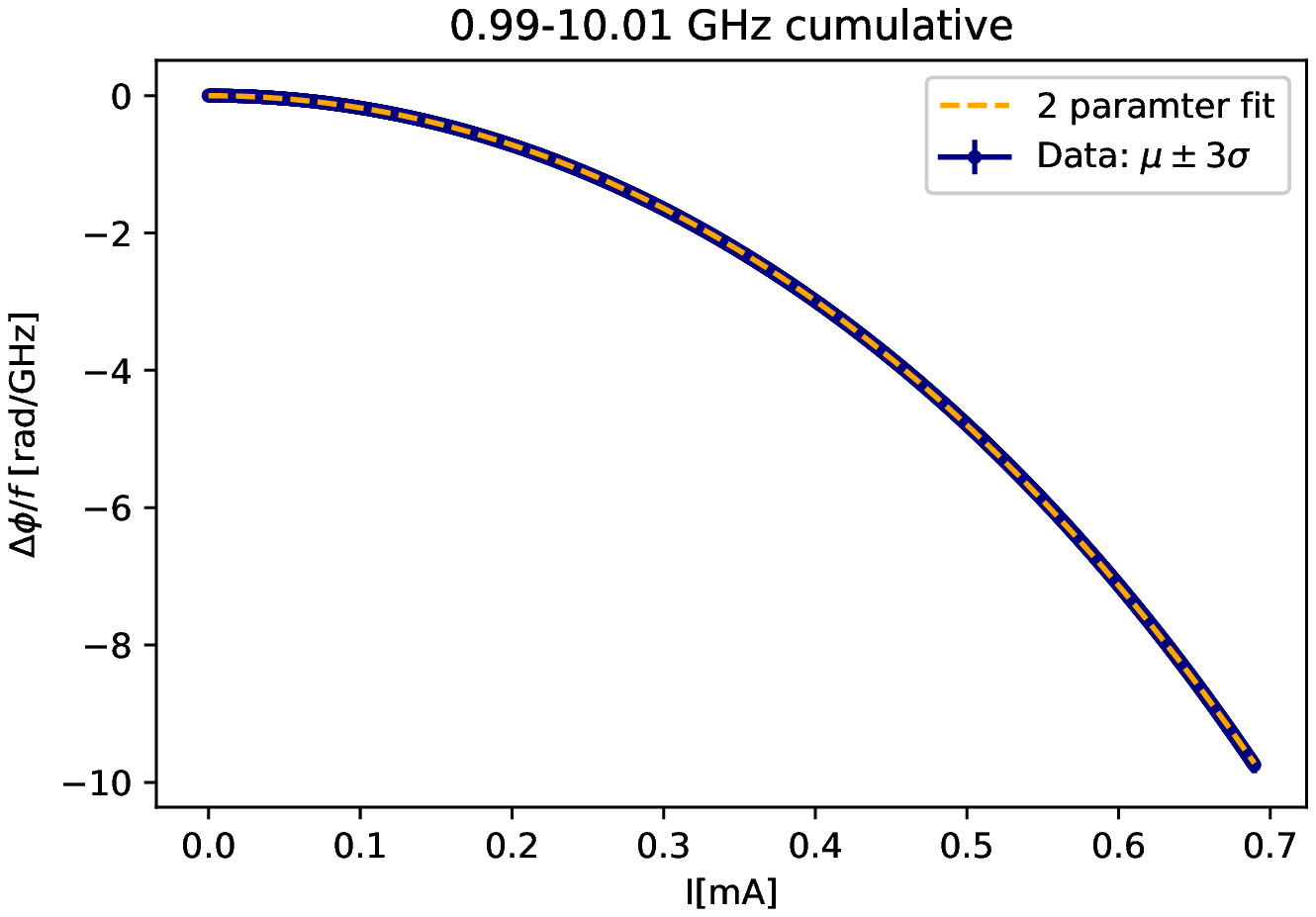}
    \textbf{Fig. 3.1} Data and fit of phase delay in STL with DC biasing.
    \label{fig:dphidf}
\end{minipage}
\begin{minipage}{.4\textwidth}
\begin{equation}
\frac{\Delta \phi (I)}{\nu} = K\left[1-\sqrt{1+\left(\frac{I}{a_I}\right)^2 +\left(\frac{I}{b_I}\right)^4}\right]
\label{eq:dphidf_model}
\end{equation}
\end{minipage}

 Fig.~\ref{fig:dphidf} shows that with 0.7 mA current\footnote{Maximum voltage applied was 1.84V via 2.66 k$\Omega$ resistor.} we obtain phase shifts $\sim\mathcal{O}(10)$ radians/GHz. Ideally $ 4\times 2\pi = 25$ radians of phase-delay is sufficient to resolve one wavelength. In this device, for 30 GHz (the lowest frequency of interest for sub-mm science) we will obtain 300 radians. This implies that for such science as Line Intensity Mapping and Cosmic Microwave Background, we can shorten our STLs from 24.7mm (chip length) to $< 5$ mm, and fabricate densely packed pixels. Broad-band antennas and low noise detectors to complete a focal plane of such SOFTS is now standard, and ever improving technology.

\section{Demonstration of a Superconducting On-chip FTS}
\vspace{-0.1cm}
\subsection{Interferometric setup and data}
\begin{figure}[!h]
	\centering
	\begin{minipage}[t]{4cm}
		\centering
		\includegraphics[scale=0.55]{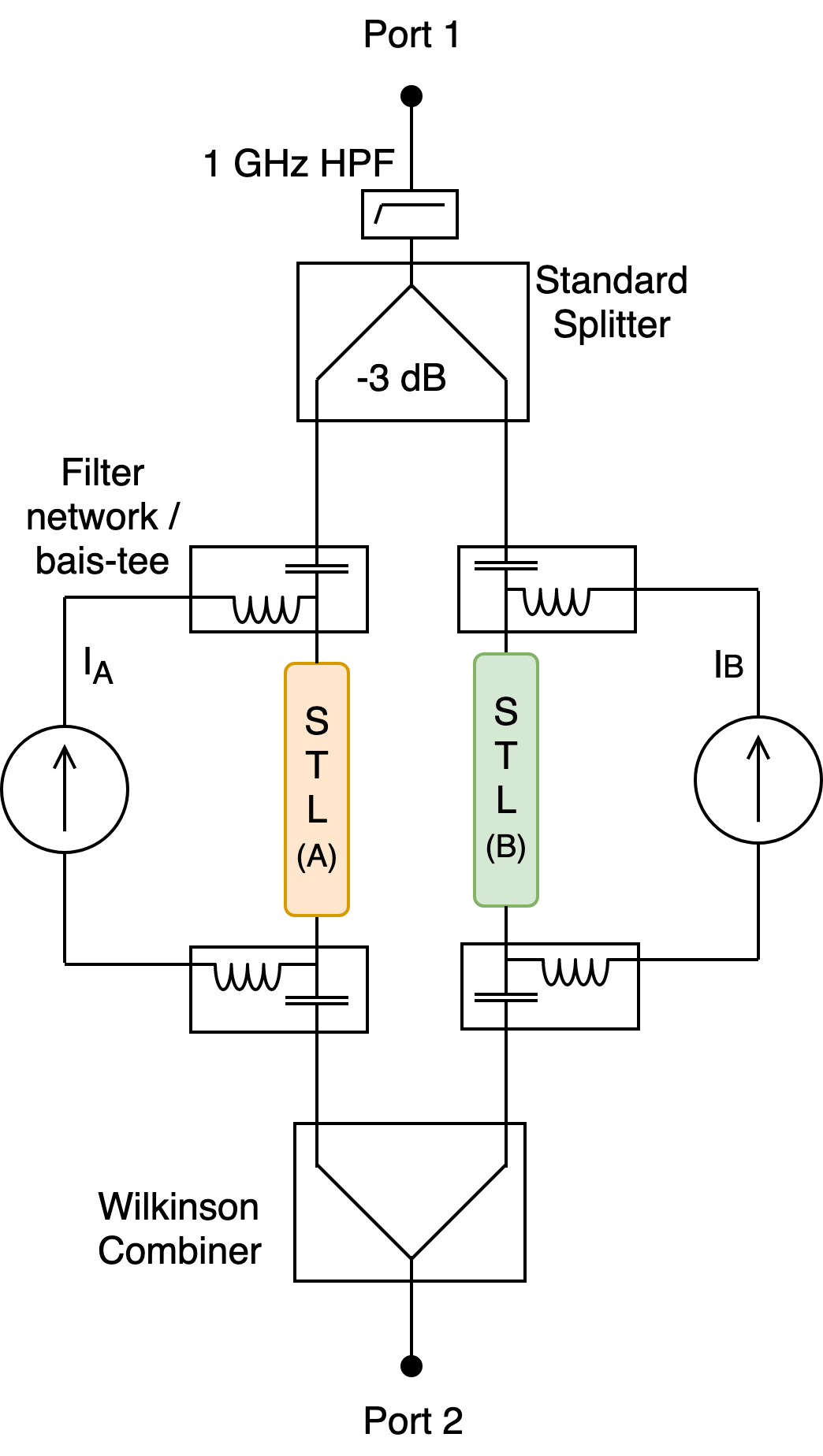}
		\caption{Experimental setup.}
		\label{fig:SOFTSJPLsetup}
	\end{minipage}
	\hspace{1.0cm}
	\begin{minipage}[t]{4cm}
		\centering
		\includegraphics[scale=0.25]{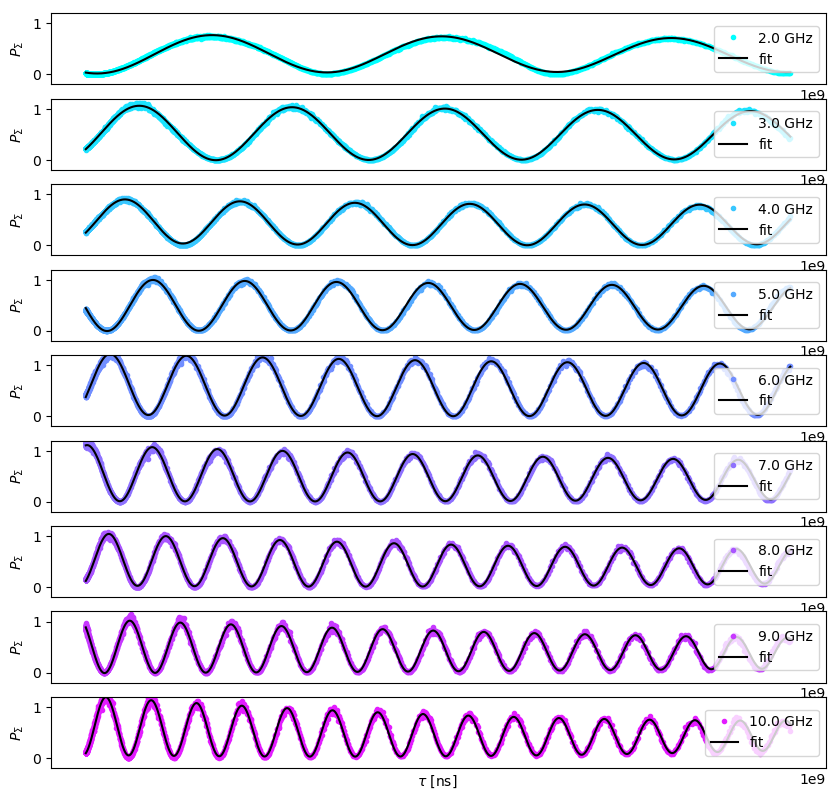}
		\caption{Summed power as function of delay,$P_{\Sigma}(\tau)$: data and fits to data for frequencies from 2-10 GHz.}
		\label{fig:SOFTSJPLdata1}
	\end{minipage}
\end{figure}

\vspace{-0.2cm}
The interferometric setup to realize one SOFTS device is shown in Fig.~\ref{fig:SOFTSJPLsetup}. We use a splitter to halve the input signal (Port 1, 100 pW, high pass filtered $>$ 1 GHz) and distribute it to two STLs. Each STL is current biased, with a low pass filter (6 kHz, via bias-tee) whilst being capacitatively coupled to the splitter.  The signals from the two STLs are summed on a broadband Wilkinson combiner. For a single tone input on Port 1, on Port 2 we should observe an interferogram that is a cosine modulation, $\displaystyle P_{\Sigma} = \frac{1}{2}(1+ \cos (\Delta \phi))$, following canonical FTS formalism. Unlike a traditional optical FTS, we have fixed physical paths, and we change the phase velocity to introduce delay ($\tau (I) = x/(u^{-1}(I) - u^{-1}(0))$), viz. $\Delta \phi = 2 \pi \nu \tau$. Measurements at some frequencies is shown in Fig.~\ref{fig:SOFTSJPLdata1}. Details on data/ fits presented thereafter.
\vspace{-0.3cm}
\vspace{-1cm}
\subsection{Discussion on measurements}
\vspace{-0.2cm}
Fig.~\ref{fig:SOFTSJPLdata1} shows that we observe the expected cosine modulation, up to some amplitude decay (discussed in detail below). The $S_{21}$ corrected interferogram for input frequency $\nu$ is  shown in Eqn.~\ref{eq:Psigmacorr}. In the limit $|S_{21}| \rightarrow 1$, we recover the usual form of $\displaystyle \frac{1}{2}(1+ \cos (2 \pi \nu \tau))$.
\begin{equation}
	P_{\Sigma}(\tau) = \frac{1}{2}\left(1+ \frac{2|S_{21,A}(\nu)||S_{21,B}(\nu)|}{|S_{21,A}(\nu)|^2 + |S_{21,B}(\nu)|^2} \cos(2\pi \nu \tau + \phi_0) \right)
\label{eq:Psigmacorr}
\end{equation}
We scale the interferograms appropriately by the individually measured $S_{21}$s of each STL, following Eqn.~\ref{eq:Psigmacorr}. During the interferometric data collection, the voltages were increased with time, and over this time temperatures increased monotonically by 0.5K, thus we expect some unbalanced profiles, or ``decay'' as seen in Fig.~\ref{fig:SOFTSJPLdata1}, particularly at higher frequencies. To analyze these data we therefore fit with a decaying cosine modulation, and compare the extracted frequencies to the input from the VNA. 

\begin{figure}[!h]
\begin{minipage}[b]{0.55\linewidth}
Our frequency fitting error is $\approx$1 MHz, and the distribution of the difference in input frequency and fit frequency has a standard deviation of 5 MHz, shown in Fig.~\ref{fig:SOFTSJPLdata1}. The expected resolution is $\Delta \nu = \frac{\delta \phi}{2 \pi} \frac{1}{\textrm{max}(\Delta \tau)}$. It is the inverse of the maximum delay, scaled with the minimum phase $\delta \phi$ we can resolve. We measure $\delta \phi \lesssim 0.04$ radians,  $\textrm{max}(\Delta \tau) \approx 1.5$ ns and we therefore expect $\Delta \nu \approx $ 4.2 MHz.
\end{minipage}
\hspace{0.2cm}
\begin{minipage}[b]{0.4\linewidth}
\centering
\includegraphics[height=9\baselineskip]{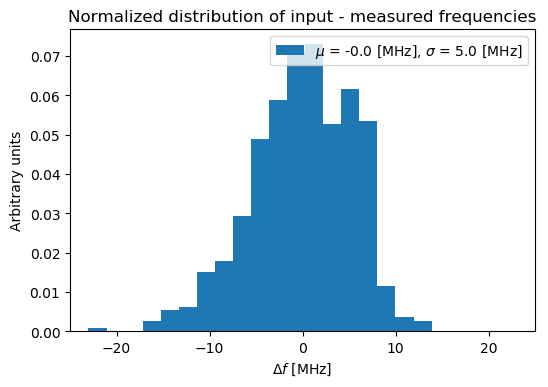}
\label{fig:SOFTSJPLdata2}
\caption{Distribution of the input minus fit frequencies.}
\end{minipage}
\end{figure}

\vspace{-0.5cm}
\section{Conclusion}
\begin{figure}[h!]
  \centering
    \includegraphics[width=0.8\textwidth]{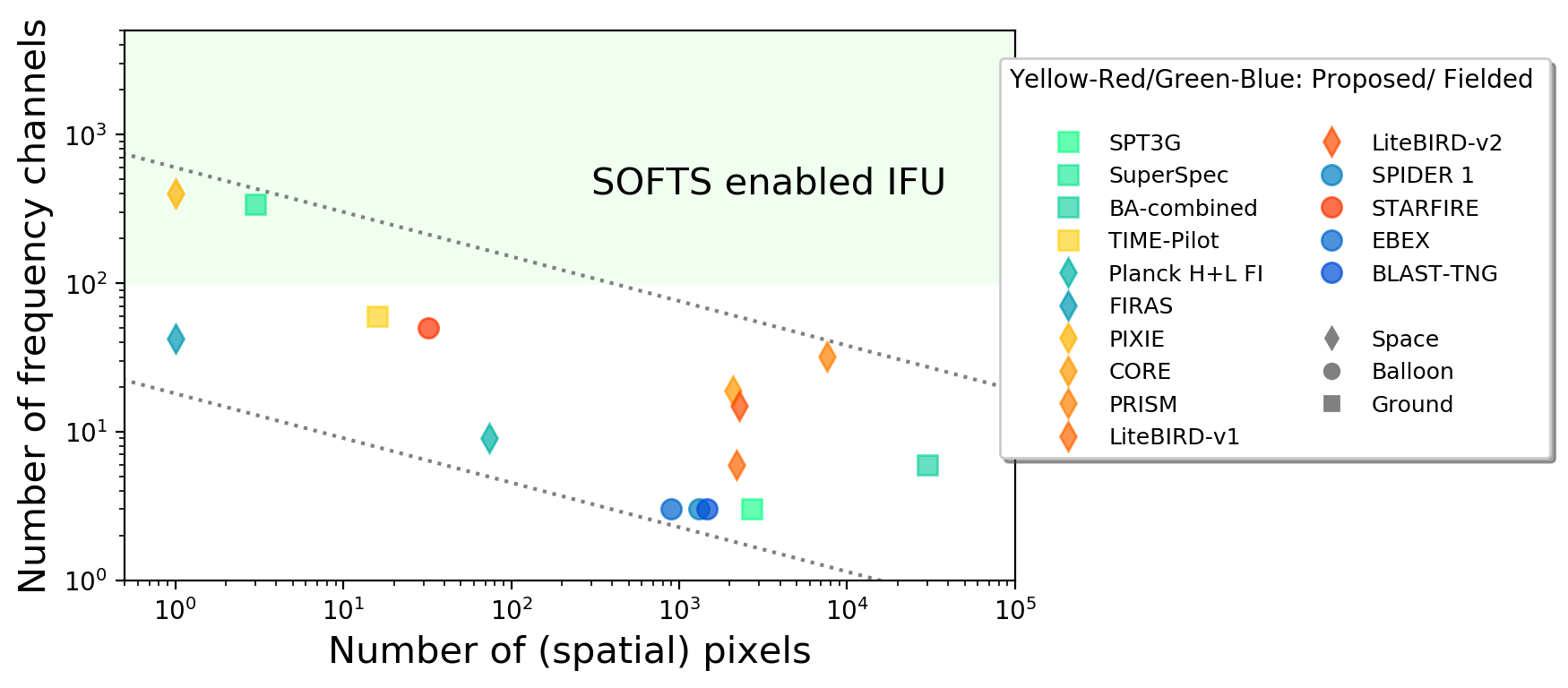}
  \caption{\small{A broad survey of mm/submm experiments expressed as number of frequency channels versus number of spatial pixels where a clear trade-off is seen,~\cite{SPIE3G, SuperSpec, BA, TIME, Planck2015, COBEFIRAS, PIXIE, CORE, PRISM, Litebird2018, LB2, SPIDER, EBEX, BLAST-TNG}. SOFTS  technology can potentially break the trade-off, enabling $>10^3$ spectral channels for all spatial pixels (shaded region).}}
  \label{fig:Npix}
\end{figure}

\vspace{-0.2cm}
We demonstrate a novel application of current controlled phase velocity modulation with thin film NbTiN superconducting transmission lines. A pair of these lines were used to make a Superconducting On-chip Fourier Transform Spectrometers (SOFTS). Our SOFTS device has a physical length-scale of 25mm, and a resolution of 5 MHz. For an optical FTS with such specifications, the physical size would have to be $\gtrsim$15 meters. We show that this on-chip device has more than sufficient delay for sub-mm science.  Our measurements indicate that individual cm-scale pixels can be fabricated; where each pixel is a SOFTS. Thus with standard broad-band antennas and low-noise detectors (TES / MKID) a kilopixel integral field unit (IFU) for sub-mm cosmology and astronomy is realizable, Fig.~\ref{fig:Npix} schematically shows the promise of SOFTS enabled IFUs, trade off between number of imaging pixels and number of spectral channels may be eliminated.
\vspace{-0.1cm}
\vspace{-0.5cm}
\begin{acknowledgements}
We thank Caltech/JPL Professors J. Bock, S. Golwala, T.C. Chang and O. Dore for helpful discussions in the early stages of this project, and H. G. LeDuc and JPL staff for device fabrication.
\end{acknowledgements}
\vspace{-0.7cm}
\bibliographystyle{ieeetr}
\bibliography{JLTP_SI_template_LTD18}

\begin{thebibliography}{10}

\bibitem{Parmenter}
R.~Parmenter, ``Nonlinear electrodynamics of super-conductors with a very small
  coherence distance,'' {\em RCA (Radio Corporation of America) Review (U.S.)},
  vol.~23, 9 1962.

\bibitem{Prober}
A.~J.~A. Prober, D.~F. Santavicca, L.~Frunzio, G.~Catelani, M.~J. Rooks,
  A.~Frydman, and D.~E, ``{Tunable superconducting nanoinductors},'' {\em
  Nanotechnology}, vol.~21, no.~44, p.~445202, 2010.

\bibitem{DayNature}
B.~{Ho Eom}, P.~K. {Day}, H.~G. {Leduc}, and J.~{Zmuidzinas}, ``{A wideband,
  low-noise superconducting amplifier with high dynamic range},'' {\em Nature
  Physics}, vol.~8, pp.~623--627, Aug. 2012.

\bibitem{Kher}
A.~Kher, {\em Superconducting Nonlinear Kinetic Inductance Devices}.
\newblock PhD thesis, California Institute of Technology, 2017.

\bibitem{Che}
G.~{Che}, S.~{Gordon}, P.~{Day}, C.~{Groppi}, R.~{Jackson}, H.~{Mani},
  P.~{Mauskopf}, H.~{Surdi}, G.~{Trichopoulos}, and M.~{Underhill}, ``{A
  Superconducting Phase Shifter and Traveling Wave Kinetic Inductance
  Parametric Amplifier for W-Band Astronomy},'' {\em ArXiv e-prints}, Oct.
  2017.

\bibitem{STLLK}
S.~M. {Anlage}, H.~J. {Snortland}, and M.~R. {Beasley}, ``A current controlled
  variable delay superconducting transmission line,'' {\em IEEE Transactions on
  Magnetics}, vol.~25, pp.~1388--1391, March 1989.

\bibitem{Ed}
E.~Schroeder, {\em Development of Superconducting Nanowire Single Photon
  Detector Technologies for Advanced Applications}.
\newblock PhD thesis, Arizona State University, 2018.

\bibitem{PICO2019}
S.~{Hanany}, M.~{Alvarez}, E.~{Artis}, P.~{Ashton}, J.~{Aumont}, R.~{Aurlien},
  R.~{Banerji}, R.~B. {Barreiro}, J.~G. {Bartlett}, S.~{Basak}, N.~{Battaglia},
  J.~{Bock}, K.~K. {Boddy}, M.~{Bonato}, J.~{Borrill}, F.~{Bouchet},
  F.~{Boulanger}, B.~{Burkhart}, J.~{Chluba}, D.~{Chuss}, S.~E. {Clark},
  J.~{Cooperrider}, B.~P. {Crill}, G.~{De Zotti}, J.~{Delabrouille}, E.~{Di
  Valentino}, J.~{Didier}, O.~{Dor{\'e}}, H.~K. {Eriksen}, J.~{Errard},
  T.~{Essinger-Hileman}, S.~{Feeney}, J.~{Filippini}, L.~{Fissel},
  R.~{Flauger}, U.~{Fuskeland}, V.~{Gluscevic}, K.~M. {Gorski}, D.~{Green},
  B.~{Hensley}, D.~{Herranz}, J.~C. {Hill}, E.~{Hivon}, R.~{Hlo{\v{z}}ek},
  J.~{Hubmayr}, B.~R. {Johnson}, W.~{Jones}, T.~{Jones}, L.~{Knox}, A.~{Kogut},
  M.~{L{\'o}pez-Caniego}, C.~{Lawrence}, A.~{Lazarian}, Z.~{Li},
  M.~{Madhavacheril}, J.-B. {Melin}, J.~{Meyers}, C.~{Murray}, M.~{Negrello},
  G.~{Novak}, R.~{O'Brient}, C.~{Paine}, T.~{Pearson}, L.~{Pogosian},
  C.~{Pryke}, G.~{Puglisi}, M.~{Remazeilles}, G.~{Rocha}, M.~{Schmittfull},
  D.~{Scott}, P.~{Shirron}, I.~{Stephens}, B.~{Sutin}, M.~{Tomasi},
  A.~{Trangsrud}, A.~{van Engelen}, F.~{Vansyngel}, I.~K. {Wehus}, Q.~{Wen},
  S.~{Xu}, K.~{Young}, and A.~{Zonca}, ``{PICO: Probe of Inflation and Cosmic
  Origins},'' {\em arXiv e-prints}, p.~arXiv:1902.10541, Feb 2019.

\bibitem{S4tech}
M.~H. Abitbol, Z.~Ahmed, D.~Barron, R.~B. Thakur, A.~N. Bender, B.~A. Benson,
  C.~A. Bischoff, S.~A. Bryan, J.~E. Carlstrom, C.~L. Chang, D.~T. Chuss, K.~T.
  Crowley, A.~Cukierman, T.~de~Haan, M.~Dobbs, T.~Essinger-Hileman, J.~P.
  Filippini, K.~Ganga, J.~E. Gudmundsson, N.~W. Halverson, S.~Hanany, S.~W.
  Henderson, C.~A. Hill, S.-P.~P. Ho, J.~Hubmayr, K.~Irwin, O.~Jeong, B.~R.
  Johnson, S.~A. Kernasovskiy, J.~M. Kovac, A.~Kusaka, A.~T. Lee, S.~Maria,
  P.~Mauskopf, J.~J. McMahon, L.~Moncelsi, A.~W. Nadolski, J.~M. Nagy, M.~D.
  Niemack, R.~C. O'Brient, S.~Padin, S.~C. Parshley, C.~Pryke, N.~A. Roe,
  K.~Rostem, J.~Ruhl, S.~M. Simon, S.~T. Staggs, A.~Suzuki, E.~R. Switzer,
  O.~Tajima, K.~L. Thompson, P.~Timbie, G.~S. Tucker, J.~D. Vieira, A.~G.
  Vieregg, B.~Westbrook, E.~J. Wollack, K.~W. Yoon, K.~S. Young, and E.~Y.
  Young, ``Cmb-s4 technology book, first edition,'' 2017.

\bibitem{SerraDore}
P.~{Serra}, O.~{Dor{\'e}}, and G.~{Lagache}, ``{Dissecting the High-z
  Interstellar Medium through Intensity Mapping Cross-correlations},'' {\em
  \apj}, vol.~833, p.~153, Dec. 2016.

\bibitem{SPIE3G}
B.~A. Benson {\em et~al.}, ``{SPT-3G: A Next-Generation Cosmic Microwave
  Background Polarization Experiment on the South Pole Telescope},'' {\em Proc.
  SPIE Int. Soc. Opt. Eng.}, vol.~9153, p.~91531P, 2014.

\bibitem{SuperSpec}
S.~{Hailey-Dunsheath}, E.~{Shirokoff}, P.~S. {Barry}, C.~M. {Bradford},
  G.~{Chattopadhyay}, P.~{Day}, S.~{Doyle}, M.~{Hollister}, A.~{Kovacs}, H.~G.
  {LeDuc}, P.~{Mauskopf}, C.~M. {McKenney}, R.~{Monroe}, R.~{O'Brient},
  S.~{Padin}, T.~{Reck}, L.~{Swenson}, C.~E. {Tucker}, and J.~{Zmuidzinas},
  ``{Status of SuperSpec: a broadband, on-chip millimeter-wave spectrometer},''
  in {\em Millimeter, Submillimeter, and Far-Infrared Detectors and
  Instrumentation for Astronomy VII}, vol.~9153 of {\em \procspie}, p.~91530M,
  Aug. 2014.

\bibitem{BA}
J.~A. {Grayson}, P.~A.~R. {Ade}, Z.~{Ahmed}, K.~D. {Alexander}, M.~{Amiri},
  D.~{Barkats}, S.~J. {Benton}, C.~A. {Bischoff}, J.~J. {Bock}, H.~{Boenish},
  R.~{Bowens-Rubin}, I.~{Buder}, E.~{Bullock}, V.~{Buza}, J.~{Connors}, J.~P.
  {Filippini}, S.~{Fliescher}, M.~{Halpern}, S.~{Harrison}, G.~C. {Hilton},
  V.~V. {Hristov}, H.~{Hui}, K.~D. {Irwin}, J.~{Kang}, K.~S. {Karkare},
  E.~{Karpel}, S.~{Kefeli}, S.~A. {Kernasovskiy}, J.~M. {Kovac}, C.~L. {Kuo},
  E.~M. {Leitch}, M.~{Lueker}, K.~G. {Megerian}, V.~{Monticue}, T.~{Namikawa},
  C.~B. {Netterfield}, H.~T. {Nguyen}, R.~{O'Brient}, R.~W. {Ogburn},
  C.~{Pryke}, C.~D. {Reintsema}, S.~{Richter}, R.~{Schwarz}, C.~{Sorenson},
  C.~D. {Sheehy}, Z.~K. {Staniszewski}, B.~{Steinbach}, G.~P. {Teply}, K.~L.
  {Thompson}, J.~E. {Tolan}, C.~{Tucker}, A.~D. {Turner}, A.~G. {Vieregg},
  A.~{Wandui}, A.~C. {Weber}, D.~V. {Wiebe}, J.~{Willmert}, W.~L.~K. {Wu}, and
  K.~W. {Yoon}, ``{BICEP3 performance overview and planned Keck Array
  upgrade},'' in {\em Millimeter, Submillimeter, and Far-Infrared Detectors and
  Instrumentation for Astronomy VIII}, vol.~9914 of {\em \procspie}, p.~99140S,
  July 2016.

\bibitem{TIME}
A.~{Crites}, J.~{Bock}, M.~{Bradford}, B.~{Bumble}, T.-C. {Chang}, Y.-T.
  {Cheng}, A.~R. {Cooray}, S.~{Hailey-Dunsheath}, J.~{Hunacek}, C.-T. {Li},
  R.~{O'Brient}, E.~{Shirokoff}, Z.~{Staniszewski}, C.~{Shiu}, B.~{Uzgil},
  M.~B. {Zemcov}, and G.~{Sun}, ``{Measuring the Epoch of Reionization using
  [CII] Intensity Mapping with TIME-Pilot},'' in {\em American Astronomical
  Society Meeting Abstracts \#229}, vol.~229 of {\em American Astronomical
  Society Meeting Abstracts}, p.~125.01, Jan. 2017.

\bibitem{Planck2015}
J.~{Alves}, F.~{Combes}, A.~{Ferrara}, T.~{Forveille}, and S.~{Shore},
  ``{Planck 2015 results},'' {\em \aap}, vol.~594, p.~E1, Sept. 2016.

\bibitem{COBEFIRAS}
J.~C. {Mather}, E.~S. {Cheng}, D.~A. {Cottingham}, R.~E. {Eplee}, Jr., D.~J.
  {Fixsen}, T.~{Hewagama}, R.~B. {Isaacman}, K.~A. {Jensen}, S.~S. {Meyer},
  P.~D. {Noerdlinger}, S.~M. {Read}, L.~P. {Rosen}, R.~A. {Shafer}, E.~L.
  {Wright}, C.~L. {Bennett}, N.~W. {Boggess}, M.~G. {Hauser}, T.~{Kelsall},
  S.~H. {Moseley}, Jr., R.~F. {Silverberg}, G.~F. {Smoot}, R.~{Weiss}, and
  D.~T. {Wilkinson}, ``{Measurement of the cosmic microwave background spectrum
  by the COBE FIRAS instrument},'' {\em \apj}, vol.~420, pp.~439--444, Jan.
  1994.

\bibitem{PIXIE}
A.~{Kogut}, J.~{Chluba}, D.~J. {Fixsen}, S.~{Meyer}, and D.~{Spergel}, ``{The
  Primordial Inflation Explorer (PIXIE)},'' in {\em Space Telescopes and
  Instrumentation 2016: Optical, Infrared, and Millimeter Wave}, vol.~9904 of
  {\em \procspie}, p.~99040W, July 2016.

\bibitem{CORE}
P.~{de Bernardis}, P.~A.~R. {Ade}, J.~J.~A. {Baselmans}, E.~S. {Battistelli},
  A.~{Benoit}, M.~{Bersanelli}, A.~{Bideaud}, M.~{Calvo}, F.~J. {Casas},
  G.~{Castellano}, A.~{Catalano}, I.~{Charles}, I.~{Colantoni}, F.~{Columbro},
  A.~{Coppolecchia}, M.~{Crook}, G.~{D'Alessandro}, M.~{De Petris},
  J.~{Delabrouille}, S.~{Doyle}, C.~{Franceschet}, A.~{Gomez}, J.~{Goupy},
  S.~{Hanany}, M.~{Hills}, L.~{Lamagna}, J.~{Macias-Perez}, B.~{Maffei},
  S.~{Martin}, E.~{Martinez-Gonzalez}, S.~{Masi}, D.~{McCarthy}, A.~{Mennella},
  A.~{Monfardini}, F.~{Noviello}, A.~{Paiella}, F.~{Piacentini}, M.~{Piat},
  G.~{Pisano}, G.~{Signorelli}, C.~Y. {Tan}, A.~{Tartari}, N.~{Trappe},
  S.~{Triqueneaux}, C.~{Tucker}, G.~{Vermeulen}, K.~{Young}, M.~{Zannoni},
  A.~{Ach{\'u}carro}, R.~{Allison}, M.~{Ashdown}, M.~{Ballardini}, A.~J.
  {Banday}, R.~{Banerji}, J.~{Bartlett}, N.~{Bartolo}, S.~{Basak},
  A.~{Bonaldi}, M.~{Bonato}, J.~{Borrill}, F.~{Bouchet}, F.~{Boulanger},
  T.~{Brinckmann}, M.~{Bucher}, C.~{Burigana}, A.~{Buzzelli}, Z.~Y. {Cai},
  C.~S. {Carvalho}, A.~{Challinor}, J.~{Chluba}, S.~{Clesse}, G.~{De Gasperis},
  G.~{De Zotti}, E.~{Di Valentino}, J.~M. {Diego}, J.~{Errard}, S.~{Feeney},
  R.~{Fernandez-Cobos}, F.~{Finelli}, F.~{Forastieri}, S.~{Galli},
  R.~{G{\'e}nova-Santos}, M.~{Gerbino}, J.~{Gonz{\'a}lez-Nuevo}, S.~{Hagstotz},
  J.~{Greenslade}, W.~{Handley}, C.~{Hern{\'a}ndez-Monteagudo},
  C.~{Hervias-Caimapo}, E.~{Hivon}, K.~{Kiiveri}, T.~{Kisner}, T.~{Kitching},
  M.~{Kunz}, H.~{Kurki-Suonio}, A.~{Lasenby}, M.~{Lattanzi}, J.~{Lesgourgues},
  A.~{Lewis}, M.~{Liguori}, V.~{Lindholm}, G.~{Luzzi}, C.~J.~A.~P. {Martins},
  A.~{Melchiorri}, J.~B. {Melin}, D.~{Molinari}, P.~{Natoli}, M.~{Negrello},
  A.~{Notari}, D.~{Paoletti}, G.~{Patanchon}, L.~{Polastri}, G.~{Polenta},
  A.~{Pollo}, V.~{Poulin}, M.~{Quartin}, M.~{Remazeilles}, M.~{Roman}, J.~A.
  {Rubi{\~n}o-Mart{\'{\i}}n}, L.~{Salvati}, M.~{Tomasi}, D.~{Tramonte},
  T.~{Trombetti}, J.~{V{\"a}liviita}, R.~{Van de Weijgaert}, B.~{van Tent},
  V.~{Vennin}, P.~{Vielva}, N.~{Vittorio}, and {for the CORE collaboration},
  ``{Exploring Cosmic Origins with CORE: The Instrument},'' {\em ArXiv
  e-prints}, May 2017.

\bibitem{PRISM}
P.~{Andr{\'e}}, C.~{Baccigalupi}, A.~{Banday}, D.~{Barbosa}, B.~{Barreiro},
  J.~{Bartlett}, N.~{Bartolo}, E.~{Battistelli}, R.~{Battye}, G.~{Bendo},
  A.~{Beno{\^i}t}, J.-P. {Bernard}, M.~{Bersanelli}, M.~{B{\'e}thermin},
  P.~{Bielewicz}, A.~{Bonaldi}, F.~{Bouchet}, F.~{Boulanger}, J.~{Brand},
  M.~{Bucher}, C.~{Burigana}, Z.-Y. {Cai}, P.~{Camus}, F.~{Casas},
  V.~{Casasola}, G.~{Castex}, A.~{Challinor}, J.~{Chluba}, G.~{Chon},
  S.~{Colafrancesco}, B.~{Comis}, F.~{Cuttaia}, G.~{D'Alessandro}, A.~{Da
  Silva}, R.~{Davis}, M.~{de Avillez}, P.~{de Bernardis}, M.~{de Petris},
  A.~{de Rosa}, G.~{de Zotti}, J.~{Delabrouille}, F.-X. {D{\'e}sert},
  C.~{Dickinson}, J.~M. {Diego}, J.~{Dunkley}, T.~{En{\ss}lin}, J.~{Errard},
  E.~{Falgarone}, P.~{Ferreira}, K.~{Ferri{\`e}re}, F.~{Finelli},
  A.~{Fletcher}, P.~{Fosalba}, G.~{Fuller}, S.~{Galli}, K.~{Ganga},
  J.~{Garc{\'{\i}}a-Bellido}, A.~{Ghribi}, M.~{Giard}, Y.~{Giraud-H{\'e}raud},
  J.~{Gonzalez-Nuevo}, K.~{Grainge}, A.~{Gruppuso}, A.~{Hall}, J.-C.
  {Hamilton}, M.~{Haverkorn}, C.~{Hernandez-Monteagudo}, D.~{Herranz},
  M.~{Jackson}, A.~{Jaffe}, R.~{Khatri}, M.~{Kunz}, L.~{Lamagna},
  M.~{Lattanzi}, P.~{Leahy}, J.~{Lesgourgues}, M.~{Liguori}, E.~{Liuzzo},
  M.~{Lopez-Caniego}, J.~{Macias-Perez}, B.~{Maffei}, D.~{Maino},
  A.~{Mangilli}, E.~{Martinez-Gonzalez}, C.~J.~A.~P. {Martins}, S.~{Masi},
  M.~{Massardi}, S.~{Matarrese}, A.~{Melchiorri}, J.-B. {Melin}, A.~{Mennella},
  A.~{Mignano}, M.-A. {Miville-Desch{\^e}nes}, A.~{Monfardini}, A.~{Murphy},
  P.~{Naselsky}, F.~{Nati}, P.~{Natoli}, M.~{Negrello}, F.~{Noviello},
  C.~{O'Sullivan}, F.~{Paci}, L.~{Pagano}, R.~{Paladino},
  N.~{Palanque-Delabrouille}, D.~{Paoletti}, H.~{Peiris}, F.~{Perrotta},
  F.~{Piacentini}, M.~{Piat}, L.~{Piccirillo}, G.~{Pisano}, G.~{Polenta},
  A.~{Pollo}, N.~{Ponthieu}, M.~{Remazeilles}, S.~{Ricciardi}, M.~{Roman},
  C.~{Rosset}, J.-A. {Rubino-Martin}, M.~{Salatino}, A.~{Schillaci},
  P.~{Shellard}, J.~{Silk}, A.~{Starobinsky}, R.~{Stompor}, R.~{Sunyaev},
  A.~{Tartari}, L.~{Terenzi}, L.~{Toffolatti}, M.~{Tomasi}, N.~{Trappe},
  M.~{Tristram}, T.~{Trombetti}, M.~{Tucci}, R.~{Van de Weijgaert}, B.~{Van
  Tent}, L.~{Verde}, P.~{Vielva}, B.~{Wandelt}, R.~{Watson}, and
  S.~{Withington}, ``{PRISM (Polarized Radiation Imaging and Spectroscopy
  Mission): an extended white paper},'' {\em \jcap}, vol.~2, p.~006, Feb. 2014.

\bibitem{Litebird2018}
A.~{Suzuki}, P.~A.~R. {Ade}, Y.~{Akiba}, D.~{Alonso}, K.~{Arnold}, J.~{Aumont},
  C.~{Baccigalupi}, D.~{Barron}, S.~{Basak}, S.~{Beckman}, J.~{Borrill},
  F.~{Boulanger}, M.~{Bucher}, E.~{Calabrese}, Y.~{Chinone}, S.~{Cho},
  A.~{Cukierman}, D.~W. {Curtis}, T.~{de Haan}, M.~{Dobbs}, A.~{Dominjon},
  T.~{Dotani}, L.~{Duband}, A.~{Ducout}, J.~{Dunkley}, J.~M. {Duval},
  T.~{Elleflot}, H.~K. {Eriksen}, J.~{Errard}, J.~{Fischer}, T.~{Fujino},
  T.~{Funaki}, U.~{Fuskeland}, K.~{Ganga}, N.~{Goeckner-Wald}, J.~{Grain},
  N.~W. {Halverson}, T.~{Hamada}, T.~{Hasebe}, M.~{Hasegawa}, K.~{Hattori},
  M.~{Hattori}, L.~{Hayes}, M.~{Hazumi}, N.~{Hidehira}, C.~A. {Hill},
  G.~{Hilton}, J.~{Hubmayr}, K.~{Ichiki}, T.~{Iida}, H.~{Imada}, M.~{Inoue},
  Y.~{Inoue}, K.~{D.}, H.~{Ishino}, O.~{Jeong}, H.~{Kanai}, D.~{Kaneko},
  S.~{Kashima}, N.~{Katayama}, T.~{Kawasaki}, S.~A. {Kernasovskiy},
  R.~{Keskitalo}, A.~{Kibayashi}, Y.~{Kida}, K.~{Kimura}, T.~{Kisner},
  K.~{Kohri}, E.~{Komatsu}, K.~{Komatsu}, C.~L. {Kuo}, N.~A. {Kurinsky},
  A.~{Kusaka}, A.~{Lazarian}, A.~T. {Lee}, D.~{Li}, E.~{Linder}, B.~{Maffei},
  A.~{Mangilli}, M.~{Maki}, T.~{Matsumura}, S.~{Matsuura}, D.~{Meilhan},
  S.~{Mima}, Y.~{Minami}, K.~{Mitsuda}, L.~{Montier}, M.~{Nagai},
  T.~{Nagasaki}, R.~{Nagata}, M.~{Nakajima}, S.~{Nakamura}, T.~{Namikawa},
  M.~{Naruse}, H.~{Nishino}, T.~{Nitta}, T.~{Noguchi}, H.~{Ogawa}, S.~{Oguri},
  N.~{Okada}, A.~{Okamoto}, T.~{Okamura}, C.~{Otani}, G.~{Patanchon},
  G.~{Pisano}, G.~{Rebeiz}, M.~{Remazeilles}, P.~L. {Richards}, S.~{Sakai},
  Y.~{Sakurai}, Y.~{Sato}, N.~{Sato}, M.~{Sawada}, Y.~{Segawa}, Y.~{Sekimoto},
  U.~{Seljak}, B.~D. {Sherwin}, T.~{Shimizu}, K.~{Shinozaki}, R.~{Stompor},
  H.~{Sugai}, H.~{Sugita}, J.~{Suzuki}, O.~{Tajima}, S.~{Takada}, R.~{Takaku},
  S.~{Takakura}, S.~{Takatori}, D.~{Tanabe}, E.~{Taylor}, K.~L. {Thompson},
  B.~{Thorne}, T.~{Tomaru}, T.~{Tomida}, N.~{Tomita}, M.~{Tristram},
  C.~{Tucker}, P.~{Turin}, M.~{Tsujimoto}, S.~{Uozumi}, S.~{Utsunomiya},
  Y.~{Uzawa}, F.~{Vansyngel}, I.~K. {Wehus}, B.~{Westbrook}, M.~{Willer},
  N.~{Whitehorn}, Y.~{Yamada}, R.~{Yamamoto}, N.~{Yamasaki}, T.~{Yamashita},
  and M.~{Yoshida}, ``{The LiteBIRD Satellite Mission - Sub-Kelvin
  Instrument},'' {\em ArXiv e-prints}, Jan. 2018.

\bibitem{LB2}
T.~{Matsumura}, Y.~{Akiba}, J.~{Borrill}, Y.~{Chinone}, M.~{Dobbs}, H.~{Fuke},
  M.~{Hasegawa}, K.~{Hattori}, M.~{Hattori}, M.~{Hazumi}, W.~{Holzapfel},
  Y.~{Hori}, J.~{Inatani}, M.~{Inoue}, Y.~{Inoue}, K.~{Ishidoshiro},
  H.~{Ishino}, H.~{Ishitsuka}, K.~{Karatsu}, S.~{Kashima}, N.~{Katayama},
  I.~{Kawano}, A.~{Kibayashi}, Y.~{Kibe}, K.~{Kimura}, N.~{Kimura},
  E.~{Komatsu}, M.~{Kozu}, K.~{Koga}, A.~{Lee}, H.~{Matsuhara}, S.~{Mima},
  K.~{Mitsuda}, K.~{Mizukami}, H.~{Morii}, T.~{Morishima}, M.~{Nagai},
  R.~{Nagata}, S.~{Nakamura}, M.~{Naruse}, T.~{Namikawa}, K.~{Natsume},
  T.~{Nishibori}, K.~{Nishijo}, H.~{Nishino}, A.~{Noda}, T.~{Noguchi},
  H.~{Ogawa}, S.~{Oguri}, I.~S. {Ohta}, N.~{Okada}, C.~{Otani}, P.~{Richards},
  S.~{Sakai}, N.~{Sato}, Y.~{Sato}, Y.~{Segawa}, Y.~{Sekimoto}, K.~{Shinozaki},
  H.~{Sugita}, A.~{Suzuki}, T.~{Suzuki}, O.~{Tajima}, S.~{Takada},
  S.~{Takakura}, Y.~{Takei}, T.~{Tomaru}, Y.~{Uzawa}, T.~{Wada}, H.~{Watanabe},
  Y.~{Yamada}, H.~{Yamaguchi}, N.~{Yamasaki}, M.~{Yoshida}, T.~{Yoshida}, and
  K.~{Yotsumoto}, ``{LiteBIRD: mission overview and design tradeoffs},'' in
  {\em Space Telescopes and Instrumentation 2014: Optical, Infrared, and
  Millimeter Wave}, vol.~9143 of {\em \procspie}, p.~91431F, Aug. 2014.

\bibitem{SPIDER}
J.~P. {Filippini}, P.~A.~R. {Ade}, M.~{Amiri}, S.~J. {Benton}, R.~{Bihary},
  J.~J. {Bock}, J.~R. {Bond}, J.~A. {Bonetti}, S.~A. {Bryan}, B.~{Burger},
  H.~C. {Chiang}, C.~R. {Contaldi}, B.~P. {Crill}, O.~{Dor{\'e}}, M.~{Farhang},
  L.~M. {Fissel}, N.~N. {Gandilo}, S.~R. {Golwala}, J.~E. {Gudmundsson},
  M.~{Halpern}, M.~{Hasselfield}, G.~{Hilton}, W.~{Holmes}, V.~V. {Hristov},
  K.~D. {Irwin}, W.~C. {Jones}, C.~L. {Kuo}, C.~J. {MacTavish}, P.~V. {Mason},
  T.~E. {Montroy}, T.~A. {Morford}, C.~B. {Netterfield}, D.~T. {O'Dea}, A.~S.
  {Rahlin}, C.~D. {Reintsema}, J.~E. {Ruhl}, M.~C. {Runyan}, M.~A. {Schenker},
  J.~A. {Shariff}, J.~D. {Soler}, A.~{Trangsrud}, C.~{Tucker}, R.~S. {Tucker},
  and A.~D. {Turner}, ``{SPIDER: a balloon-borne CMB polarimeter for large
  angular scales},'' in {\em Millimeter, Submillimeter, and Far-Infrared
  Detectors and Instrumentation for Astronomy V}, vol.~7741 of {\em \procspie},
  p.~77411N, July 2010.

\bibitem{EBEX}
{The EBEX Collaboration}, M.~{Abitbol}, A.~M. {Aboobaker}, P.~{Ade},
  D.~{Araujo}, F.~{Aubin}, C.~{Baccigalupi}, C.~{Bao}, D.~{Chapman},
  J.~{Didier}, M.~{Dobbs}, S.~M. {Feeney}, C.~{Geach}, W.~{Grainger},
  S.~{Hanany}, K.~{Helson}, S.~{Hillbrand}, G.~{Hilton}, J.~{Hubmayr},
  K.~{Irwin}, A.~{Jaffe}, B.~{Johnson}, T.~{Jones}, J.~{Klein}, A.~{Korotkov},
  A.~{Lee}, L.~{Levinson}, M.~{Limon}, K.~{MacDermid}, A.~D. {Miller},
  M.~{Milligan}, K.~{Raach}, B.~{Reichborn-Kjennerud}, C.~{Reintsema},
  I.~{Sagiv}, G.~{Smecher}, G.~S. {Tucker}, B.~{Westbrook}, K.~{Young}, and
  K.~{Zilic}, ``{The EBEX Balloon Borne Experiment - Detectors and Readout},''
  {\em ArXiv e-prints}, Mar. 2018.

\bibitem{BLAST-TNG}
L.~M. {Fissel}, P.~{Ade}, F.~E. {Angil{\`e}}, P.~{Campbell Ashton}, J.~E.
  {Austermann}, T.~{Billings}, G.~{Che}, H.-M. {Cho}, M.~R. {Cunningham},
  K.~{Davis}, M.~J. {Devlin}, S.~{Dicker}, B.~{Dober}, Y.~{Fukui},
  N.~{Galitzki}, j.~{gao}, S.~{Gordon}, C.~E. {Groppi}, S.~{Hillbrand},
  G.~{Hilton}, H.~{Hubmayr}, K.~{Irwin}, P.~{Jones}, J.~{Klein}, d.~{li}, Z.-Y.
  {Li}, n.~{lourie}, I.~{Lowe}, H.~{Mani}, P.~G. {Martin}, P.~{Mauskopf},
  C.~{McKenney}, F.~{Nati}, G.~{Novak}, E.~{Pascale}, g.~{pisano}, F.~{Pereira
  Santos}, D.~{Scott}, A.~{Sinclair}, J.~{Diego Diego Soler}, c.~{tucker},
  M.~{Underhill}, M.~{Vissers}, and P.~{Williams}, ``{BLAST-TNG: A Next
  Generation Balloon-borne Large Aperture Submillimeter Polarimeter},'' in {\em
  American Astronomical Society Meeting Abstracts \#229}, vol.~229 of {\em
  American Astronomical Society Meeting Abstracts}, p.~133.06, Jan. 2017.

\end{thebibliography}

\end{document}